# Inconsistency of Carnot's theorem's proof by William Thomson

## V. Ihnatovych

Department of Philosophy, National Technical University of Ukraine "Kyiv Polytechnic Institute",
Kyiv, Ukraine
e-mail: V.Ihnatovych@kpi.ua

**Abstract**

William Thomson proved Carnot's theorem basing on postulate: "It is impossible, by means of inanimate material agency, to derive mechanical effect from any portion of matter by cooling it below the temperature of the coldest of the surrounding objects". The present paper demonstrates that Carnot's theorem can be proved based on the contrary Thomson's postulate: "It is impossible to use the mechanical effect to the heating the coldest of surrounding objects". A conclusion that Carnot's theorem does not follow from the Thomson's postulate has been drawn.

### 1. Introduction

The Carnot's heat engine is described almost in all courses of physics and thermodynamics (see for example [2, 6–9]).

Carnot engine operates on the reversible Carnot cycle. The direct Carnot's cycle includes four steps: 1) isothermal enlargement of working substance at temperature $T_1$, which equals to the temperature of hotter reservoir of heat; 2) adiabatic enlargement of working substance, in the process of which it's temperature lowers to the temperature $T_2$, which equals to the temperature of colder reservoir of heat; 3) isothermal compression of working substance at temperature $T_2$; 4) adiabatic compression of working substance, in the process of which it's temperature rises to the temperature $T_1$.

In the direct cycle the Carnot's engine gets from the hotter reservoir for one cycle the amount of heat $Q_1$, transfers the amount of heat $Q_2$ ($Q_2 < Q_1$) to the colder reservoir, and the difference between $Q_1$–$Q_2$ converts to the work $L$, which can be used to lifting of weight.

In the reverse cycle the Carnot's engine gets from the colder reservoir for one cycle the amount of heat $Q_2$, transfers the amount of the heat $Q_1$ to the hotter reservoir and converts into heat work $L$, obtained, for example, by of the lowering of a weight.

$Q_1 = Q_2 + L$.

The efficiency of a heat engine (cycle efficiency) η is defined by:

$$\eta = \frac{L}{Q_1} = \frac{Q_1 - Q_2}{Q_1}$$

By Carnot's theorem efficiency of the Carnot's heat engine is independent of the working substance used in the engine (see for example [2, p.29-35; 6, p.69-72; 7, p.63-65; 8, p.97-99; 9, p.27-29]).

William Thomson gave in the work "On the Dynamical Theory of Heat" the following statement of the Carnot's theorem: "If an engine be such that, when it is worked backwards, the physical and mechanical agencies in every part of its motions are all reversed, it produces as much mechanical effect as can be produced by any thermodynamic engine, with the same temperatures of source and refrigerator, from a given quantity of heat" [10, p.178].

Above in that work he wrote: "The whole theory of the motive power of heat is founded on the two following propositions, due respectively to Joule, and to Carnot and Clausius" [10, p.178]. This theorem he called "Proposition II (Carnot and Clausius)".

Thomson proved Carnot's theorem basing on postulate: "*It is impossible, by means of inanimate material agency, to derive mechanical effect from any portion of matter by cooling it below the temperature of the coldest of the surrounding objects*" [10, p.179].

It is known that the Thomson's postulate is equivalent to Clausius's postulate, according to which the heat can not pass spontaneously from a colder to a warmer reservoir (see for example [2, 6, 8, 9]).

A. A. Gukhman showed that Carnot's theorem can be proved on the basis of the "antipostulate" contrary to Clausius's postulate, according to which the heat can not pass spontaneously from a warmer to a colder reservoir [3, p.79-80; 4, p.340-341] (see also [5]).

The present paper demonstrates that Carnot's theorem can be proved based on the "antipostulate" contrary to Thomson's postulate: "It is impossible to use the mechanical effect to the heating the coldest of surrounding objects".

**2. The proof of Carnot's theorem based on Thomson's postulate**

Assume there are two Carnot engine, which efficiency are $\eta_1$ and $\eta_2$, where $\eta_1 > \eta_2$. The engine *1* is running on the direct Carnot cycle, the engine *2* – the reverse Carnot cycle. The engine *1* for one cycle gets the amount of heat $Q_1$ from the hotter reservoir, transfers the amount of heat $Q_2$ to colder reservoir and produces work $L$. The engine *2* gets the amount of heat $Q_2'$ from the colder reservoir, gets from the engine *1* the amount of work $L'$, transfer to the hotter reservoir the amount of heat $Q_1$.

$Q_1 = Q_2 + L = Q_2' + L'$.

If $\eta_1 > \eta_2$, then $Q_2' > Q_2$ and $L' < L$.

The engine *1* produces work *L*. Part of this work is equal to *L'* is giving to the engine *2*, part is equal to *L* – *L'* is spending on lifting of weight *G* (see Fig. 1).

$L - L' = Q_2' - Q_2$.

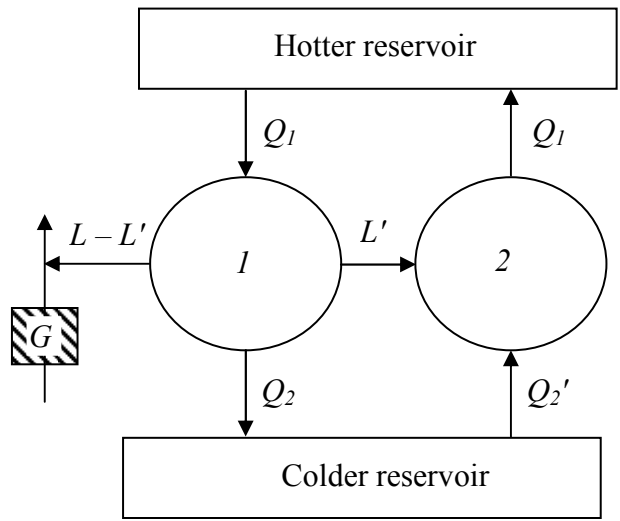

Fig. 1

The result of the two engines is the raising of a weight and the cooling the colder reservoir ($Q_2' > Q_2$). This one contradicts to the Thomson's postulate: "*It is impossible, by means of inanimate material agency, to derive mechanical effect from any portion of matter by cooling it below the temperature of the coldest of the surrounding objects*". Consequently, the assumption $\eta_1 > \eta_2$ is false.

Suggested, $\eta_1 < \eta_2$. When the engine *1* is running on reverse Carnot cycle and the engine *2* is running on the direct Carnot cycle, we get again a result that contradicts the Thomson's postulate.

Efficiency of the engine *1* can be neither less nor more efficient engine *2*.

The efficiency of the Carnot engine is independent of the working substance.

**3. The proof of the Carnot's theorem based on "antipostulate"**

Assume there are two Carnot engine, which efficiency are $\eta_1$ and $\eta_2$, where $\eta_1 < \eta_2$. The engine *1* is running on the direct Carnot cycle, the engine *2* – the reverse Carnot cycle. The engine *1* for one cycle gets the amount of heat $Q_1$ from the hotter reservoir, transfers the amount of heat $Q_2$ to colder reservoir and produces work *L*. The engine *2* gets the amount of heat $Q_2'$ from the colder reservoir, gets of the engine *1* the amount of work *L*, from an external source – the amount of work $L' - L$, transfer to the hotter reservoir the amount of heat $Q_1$.

$Q_1 = Q_2 + L = Q_2' + L'$.

If $\eta_1 < \eta_2$, then $Q_2 > Q_2'$ and $L < L'$.

The engine *2* absorbs work $L$, which is produced by engine *1*, and the work $L' - L$, which is obtained by lowering of weight $G$ (see Fig. 2).

$L' - L = Q_2 - Q_2'$.

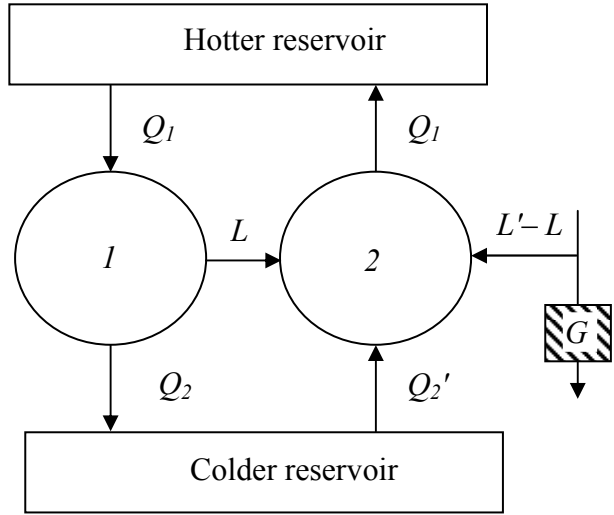

Fig. 2

The result of the two engines working is lowering of a weight and heating of colder reservoir ($Q_2 > Q_2'$). This contradicts to the "antipostulate": *"It is impossible to use the mechanical effect to the heating the coldest of surrounding objects"*. Consequently, the assumption $\eta_1 < \eta_2$ is false.

Suggested that $\eta_1 > \eta_2$. When the engine *1* is running on reverse Carnot cycle and the engine *2* is running on the direct Carnot cycle, we get again a result that contradicts the "antipostulate".

Efficiency of the engine *1* can be neither less nor more efficient engine *2*.

The efficiency of the Carnot engine is independent of the working substance.

**4. Discussion**

Same true conclusion cannot follows from two contrary assertions. Since the proof of Carnot's theorem is not changed substantially when Thomson's postulate is replaced with contrary "antipostulate", it can be concluded that Carnot's theorem is not in a logical connection with Thomson's postulate.

This result is consistent with Gukhman's derivation concerning Clausius's postulate.

However, from this one can not make conclusions about falsity of Carnot's theorem as well as Clausius's and Thomson's postulates.

**5. Conclusions**

Carnot's theorem's proof based on Thomson's postulate about impossibility. Carnot's theorem does not follow from Thomson's postulate: "It is impossible, by means of inanimate material agency, to derive mechanical effect from any portion of matter by cooling it below the temperature of the coldest of the surrounding objects". Thomson's proof of Carnot's theorem should be taken out from thermodynamics courses.